\documentclass[showpacs,prb,aps,amsmath,amssymb,floatfix,twocolumn,floats]{revtex4}
\usepackage{epsfig}
\usepackage{graphicx}
\newcommand{\ql}{q_{{\bf l}}}
\begin{document}

\title{On the valence-bond solid phase of the crossed-chain quantum spin model}
\author{Wolfram Brenig and Matthias Grzeschik}
\affiliation{Institut f\"ur Theoretische Physik,
Technische Universit\"at Braunschweig, 38106 Braunschweig,
Germany}
\date{\today}

\begin{abstract}
\begin{center}
\parbox{14cm}{
Using a series expansion based on the flow-equation method
we study the ground state energy and the elementary triplet excitations of a
generalized model of crossed spin-$1/2$ chains starting from the limit
of decoupled quadrumers. The triplet dispersion is shown to be
very sensitive to the inter-quadrumer frustration, exhibiting a line of almost
complete localization as well as lines of  quantum phase transitions limiting
the stability of the valence-bond solid phase. In the vicinity of the
checkerboard-point a finite window of exchange couplings is found with a
non-zero spin-gap, consistent with known results from exact diagonalization.
The ground state energy is lower than that of the bare quadrumer
case for all exchange couplings investigated. In the limiting situation of the
fully
frustrated checkerboard magnet our results agree with earlier series expansion
studies.}

\end{center}
\end{abstract}


\maketitle

\section{Introduction}
Frustrated quantum magnets have attracted considerable interest recently.
On the one hand this is related to the ongoing quest for systems
exhibiting spin liquid behavior and quantum disorder \cite{Lhuillier02a}.
On the other hand numerous materials displaying geometrical or quantum
chemical frustration of the magnetic exchange have been discovered recently.
Prominent examples are the one-dimensional (1D) frustrated spin-Peierls
compound CuGeO$_3$\cite{Hase93a} or the 2D orthogonal spin-dimer system
SrCu$_2$(BO$_3$)$_2$ with frustrating inter-dimer coupling\cite{Kageyama00a}
as well as the 3D tetrahedral tellurate compounds Cu$_2$Te$_2$O$_5$X$_2$,
with X=Cl, Br\cite{Lemmens01a,Brenig03a}. One promising route into
spin-liquid behavior is via the coupling of locally frustrated units like
triangles or tetrahedra, as in the kagom\'e, the pyrochlore, and the
checkerboard (i.e. planar pyrochlore) lattices. Both, on the kagom\'e and the
checkerboard lattice low-lying singlets seem to exist with no long-range
magnetic order (LRO) in the former\cite{Lecheminant97Waldtmann98,Mila00a},
and a valence-bond-crystal (VBC) ground state in the latter
case\cite{Fouet03a,Brenig02a,Canals02a,Sindzingre02a}.
The physics of the 3D pyrochlore quantum-magnet is still far from
clarified with a hope that analysis of the planar model may lead to
further progress in three 
dimensions\cite{Koga01a,Tsunetsugu01a,Tsunetsugu02a,Berg03a}.

\begin{figure}[bt]
\center{\psfig{file=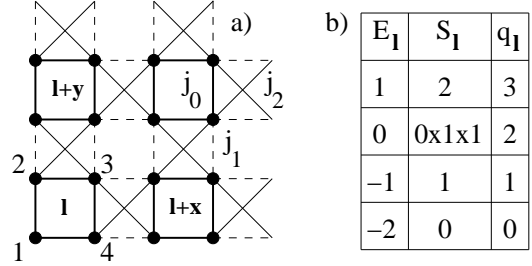,width=0.8\columnwidth}\hspace*{.6cm}}
\caption[l]{a) The generalized crossed-chain model. Spin-1/2 moments are
located on the solid circles. Thin solid(Thick solid and thin dashed) lines 
refer
to the crossed chains(inter-chain coupling). The generic model with complete
frustration of the inter chain coupling occurs at $j_0=j_1$. The quadrumer
exchange is set to $j_0=1$ hereafter with $j_1$ and $j_2$ corresponding to
the expansion parameters. b) Local energy $E_{\bf l}$, total spin $S_{\bf
l}$,  and quantum-number $\ql$  of a single quadrumer.}
\label{fig1}
\end{figure}

A closely related approach to spin-liquids above one dimensions has
become of interest recently based on the frustrated coupling of
spin-1/2 Heisenberg chains to form the planar crossed-chain model
(CCM)\cite{Emery00a,Vishwanath01a,Starykh02a}. A generalized version
of this is depicted in fig.~\ref{fig1} which resembles the generic CCM
for $j_0$=$j_1$. The generic CCM interpolates between three rather distinct
regimes. Due to the complete frustration of the inter-chain exchange
at $j_0=j_1$ and for $j_2/j_1\gtrsim 1.25$, it has been argued that
the CCM stabilizes a 2D 'sliding Luttinger liquid'
(SLL)\cite{Emery00a,Vishwanath01a,Starykh02a}. This is in contrast to
the instability of coupled Heisenberg chains
with respect to spinon-binding and the formation of antiferromagnetism
(AFM) or VBC order if linked via {\em non}-frustrating exchange. The
SLL shows no LRO with massless, deconfined spinons forming the
elementary excitations. At $j_0$=$j_1$=$j_2$ the CCM is identical to the
checkerboard magnet exhibiting the aforementioned VBC ground state
and a spin gap\cite{Fouet03a,Brenig02a,Canals02a,Sindzingre02a}. Finally,
for $j_2\to 0$ and $j_0$=$j_1$ the CCM maps onto the 2D spin-1/2
Heisenberg model on the square-lattice with AFM LRO and gapless magnon
excitations. The AFM LRO in the latter region and the VBC have been
suggested to coexist in the vicinity of the point $j_2/j_1 \sim
0.7$\cite{Sushkov01a,Sachdev02a}, which however has been questioned
based on results from exact diagonalization (ED)\cite{Sindzingre02a}.

The CCM is under intense current investigation. The SLL limit has been
analyzed by
bosonization\cite{Emery00a,Vishwanath01a,Starykh02a}. Numerical
studies have been performed on up to 36
spins\cite{Fouet03a,Brenig02a,Sindzingre02a,Palmer01a}. In particular, ED for
$j_0$=$j_1$ in ref.\cite{Sindzingre02a} suggests the VBC spin-gap to
close below $j_2/j_1\lesssim 0.65$ and moreover 1D, i.e. SLL, behavior
is found above $j_2/j_1\gtrsim 1.5$. Several analytical methods have been
employed, including various semiclassical, Sp(N), and linear-spin-wave
(LSW) approaches incorporating $1/S$-corrections
also\cite{Singh98a,Lieb99,Moessner01a,Canals02a,Tchernyshyov03a}.
LSW predicts stability of the AFM LRO for $j_2/j_1\leq 0.76$ at
$j_0$=$j_1$ and spin-1/2\cite{Singh98a}. Because it is not even obvious
that extrapolations from the large-$S$ or -$N$ limit will lead to conclusive
answers for the quantum case, it seems highly desirable to obtain
results from other analytic approaches applicable to $S$=1/2, such as
eg. series expansions (SE).  At $j_1$=$j_2$ SE has been performed
starting from the quadrumer\cite{Brenig02a} and the
tetrahedral\cite{Elhajal01a} limit.  However, SE for $j_1\neq j_2$ is not
available.

In this context it is the purpose of this work to shed more light onto the
quantum CCM by studying its elementary excitations, the stability of the
VBC phase,  and the ground state properties. This will be done by series
expansions starting from the limit of decoupled quadrumers. The paper is
organized as follows. In section two we describe our method of
calculation. In section three and four we discuss the triplet excitations
and the instabilities of the VBC phase with respect to triplet softening.
Section five is devoted to the ground state energy. A summary and some
technical details are provided in the conclusions and the appendix.

\section{Series Expansion by Continuous Unitary Transformation}
Recently it has been shown, that the checkerboard point of the CCM, i.e.
$j_1$=$j_2$=1, can be described by SE in terms of the single
coupling constant $j$=$j_1$=$j_2$, starting from the limit of decoupled
quadrumes\cite{Brenig02a}. The SE was found to adiabatically connect
the bare VBC at $j=0$ to a renormalized one at $j=1$ without being
hindered by quantum phase transitions. This is consistent with other
findings of a VBC at
$j_1$=$j_2$=1\cite{Fouet03a,Sindzingre02a,Canals02a}. Motivated by this
we decompose the Hamiltonian of the CCM in a form adapted to a VBC
symmetry breaking and normalized to the overall
unit of energy $j_0$ which will be set equal to unity hereafter
\begin{eqnarray}
\label{eq1}
H&&=\sum_{{\bf l}} \left[
\sum_{i=1...4} {\bf S}_{i{\bf l}}{\bf S}_{i+1{\bf l}}
+j_1(
{\bf S}_{2{\bf l}}{\bf S}_{1{\bf l}+{\bf y}}+
{\bf S}_{3{\bf l}}{\bf S}_{4{\bf l}+{\bf y}}+
\right. \nonumber\\
&&{\bf S}_{3{\bf l}}{\bf S}_{2{\bf l}+{\bf x}}+
{\bf S}_{4{\bf l}}{\bf S}_{1{\bf l}+{\bf x}})
+j_2(
{\bf S}_{2{\bf l}}{\bf S}_{4{\bf l}+{\bf y}}+
{\bf S}_{3{\bf l}}{\bf S}_{1{\bf l}+{\bf y}}+
\nonumber\\[8pt]
&&\left.
{\bf S}_{3{\bf l}}{\bf S}_{1{\bf l}+{\bf x}}+
{\bf S}_{4{\bf l}}{\bf S}_{2{\bf l}+{\bf x}})
\right]
\\
\label{eq2}
&&=H_0  +  \sum^{N}_{n=-N}(j_1T_{1n}+j_2T_{2n} )
\end{eqnarray}
${\bf S}_{i{\bf l}}$ refers to spin-$1/2$ operators residing on the vertices
$i=1\ldots 4$ of the quadrumer at site ${\bf l}$ with periodic boundary
conditions on $i$, c.f.\ fig.~\ref{fig1}a). $H_0$ is the sum
over local quadrumer Hamiltonians the spectrum of which, c.f.
fig.~\ref{fig1}b), consists of four equidistant levels which can be labeled
by the local total spin $S_{\bf l}$ and a number of local energy quanta $\ql$.
$H_0$ displays an equidistant ladder spectrum
labeled by $Q=\sum_{{\bf l}} \ql$. The $Q$=0 sector is the {\em unperturbed}
ground state $|0\rangle$ of $H_0$, which is a VBC
of quadrumer-singlets. The $Q$=1-sector of $H_0$ consists of linear
combinations of local $S_{\bf l}$=1 single-particle excitations of the VBC
with $\ql$=1, where ${\bf l}$ runs over the lattice. At $Q$=2 the
spectrum of $H_0$ has {\em total} spin $S$=0,1, or 2 and is of multiparticle
nature. Eg., for total $S$=0 at $Q$=2 it comprises of one-particle singlets
with $\ql$=2 and two-particle singlets constructed from triplets
with $\ql$=$q_{{\bf m}}$=1 and ${\bf l}\neq {\bf m}$. In turn, the perturbing
terms in (\ref{eq1}) $\propto j_{1,2}$, can be understood as a sum
of operators $T_{1n,2n}$ which {\em nonlocally} create(destroy) $n\geq 0$
($n<0$) quanta within the ladder spectrum of $H_0$.

It has been shown\cite{Stein97,Mielke98,Knetter00,Brenig02a,Brenig03a}
that problems of type (\ref{eq2}) allow for perturbative analysis
using a continuous unitary
transformation generated by the flow equation method of 
Wegner\cite{Wegner94}. The unitarily rotated effective
Hamiltonian $H_{\rm eff}$ reads\cite{Stein97,Knetter00}
\begin{eqnarray}
\label{w3}
H_{\rm eff}=H_0+\sum^\infty_{n=1}
\sum_{\stackrel{\mbox{\scriptsize $|{\bf m}|=n$}}{M({\bf m})=0}}
C({\bf m})\;W_{m_1}W_{m_2}\ldots W_{m_n}
\end{eqnarray}
where ${\bf m}=(m_1\ldots m_n)$ with $|{\bf m}|=n$ is an $n$-tuple of
integers, each in a range of $m_i\in\{0,\pm 1,\ldots,\pm N\}$ and
$W_{n}=j_1 T_{1n}+j_2 T_{2n}$. In contrast to $H$ of (\ref{eq1}),
$H_{\rm eff}$ {\em conserves} the total number of quanta $Q$. This
is evident from the constraint $M({\bf m})=\sum^n_{i=1} m_i=0$.
The amplitudes $C({\bf m})$ are rational numbers computed from the
flow equation method \cite{Stein97,Knetter00}. Explicit
tabulation\cite{HttpRef} of the $T_{in}$ shows that for the Hamiltonian
in (\ref{eq1}) $N=4$. In the context of spin systems the flow equation
approach has been applied successfully to 1D and 2D dimer-models
\cite{Knetter00,MH00}, as well as to 2D and 3D quadrumer-models
\cite{Brenig02a,Brenig03a}

$Q$-conservation of $H_{\rm eff}$ leads to a ground state energy of
$E_g=\langle 0|H_{\rm eff}|0\rangle$. $Q$-conservation also guarantees the
$Q$=1-triplets to remain genuine one-particle states, i.e., their dispersion
can be calculated via $E_\mu({\bf k})=\sum_{lm} t_{\mu,lm} e^{i (k_x l+k_y m)}$
where $t_{\mu,lm}=\langle \mu,lm|H_{\rm eff}|\mu,00\rangle - \delta_{lm,00}
E^{obc}_g$ are hopping matrix elements from site $(0,0)$ to site $(l,m)$ for
a quadrumer excitation $\mu$ inserted into the unperturbed ground state.
$t_{\mu,lm}$ has to be evaluated on clusters with
open boundary conditions large enough to embed all linked paths of length $n$
connecting sites $(0,0)$ to $(l,m)$ at $O(n)$ of the perturbation. $E^{obc}_g
=\langle 0|H_{\rm eff}|0\rangle$ on the $t_{\mu,00}$-cluster. States from
sectors with $Q > 1$ will not only disperse via
$H_{\rm eff}$ but require the solution of an interacting problem.

\section{Elementary Triplet Excitations}\label{TExcits}
\begin{figure}[tb]
\center{\psfig{file=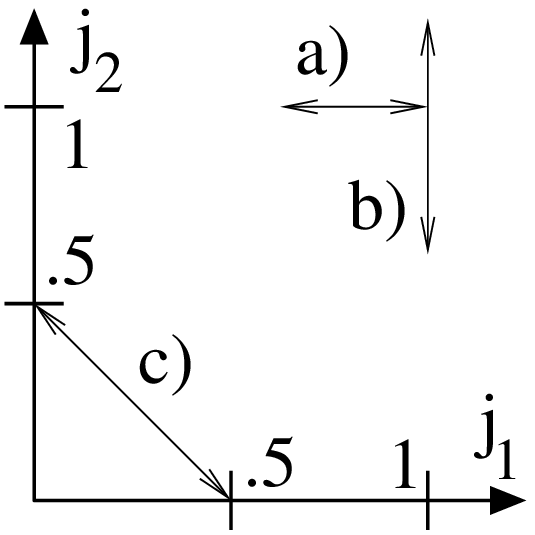,width=0.3\columnwidth}}
\center{\psfig{file=fig2b.eps,width=0.8\columnwidth}\hspace*{.5cm}}
\caption{Evolution of the triplet dispersion $E_T({\bf k},j_1, j_2)$ for
various $j_1$, $j_2$ chosen along the directions a), b), and c) sketched
in the top panel and for a path of ${\bf
k}$=($0,0$)-($\pi,0$)-($\pi,\pi$)-($0,0$) in the 2D Brillouin zone.
Solid[dashed] curves in panels a)-c)[a),b)]: $O(5)$[$O(4)$] SEs. Dots in
panel c): different $O(5)$ SE for the plaquette square-lattice from
ref.\cite{Koga99a}.}
\label{fig2}
\end{figure}

We have evaluated the one particle hopping matrix elements up to fifth
order. They are listed explicitly in the appendix. Several of the resulting
dispersions $E_T({\bf k},j_1,j_2)$ are shown in fig.~\ref{fig2} both, as a
function of the wave vector along the irreducible wedge of the Brillouin
zone of the 2D square lattice and for various values of the coupling
constants $j_1$ and $j_2$.

Two effects can be extracted from fig.~\ref{fig2}. First, it demonstrates
that the CCM 'localizes' triplet excitations on the line of largest
inter-quadrumer
frustration, i.e.  for $j_1=j_2$. This can be observed both, from the
three panels a), b), and c) where the dispersions seem completely flat for
$j_1$=$j_2$, but also from the explicit expressions for
$t_{lm}$ in appendix \ref{appB}. I.e., up to $O(4)$ and for
$l\neq m$ all hopping matrix elements are proportional to powers of
$(j_1-j_2)$. For $l\neq m$ and $lm\neq 10$ this remains true even up
to $O(5)$, where however $t_{10}\neq 0$. Therefore the
triplet-localization at maximum frustration is {\em not} complete and
a weak hopping remains beyond $O(4)$. Inserting $j_1$=$j_2$
into (\ref{b1}-\ref{b13}) we find that $E_T({\bf k},j_1=j_2)$ is exactly
identical up to $O(5)$ with an $O(7)$ SE which is available for this
case\cite{Brenig02a}.

The second fact to note from fig.~\ref{fig2} is that the VBC
ground-state is unstable with respect to triplet-softening,
i.e. magnetic ordering at suitably chosen values of the coupling
constants. Depending on the ratio of $j_2/j_1>1$ or $<1$ this
instability will occur either at a critical wave vector of ${\bf
k}_c=$($\pi,\pi$) or (0,0). Consistently with this the structure of
exchange for the CCM, although topologically distinct, is bipartite
both, for $j_1\neq 0$, $j_2\rightarrow 0$ and for $j_1\rightarrow 0$
and $j_2\neq 0$. The instabilities are studied in more detail in the next
section.

Finally, panel a) and b) of fig.~\ref{fig2} contain $O(4)$ SEs for a
comparison with the $O(5)$ results. The difference is small only. For $j_1$
and $j_2$ along path c) the difference is hardly visible and the $O(4)$ SE
has not been displayed in the corresponding panel. The case of $j_1=0.5$
and $j_2=0$ in panel c) is identical to a particular realization of the
plaquette square-lattice Heisenberg magnet. The triplet dispersion of this
system has been investigated with a rather different type of
fifth-order plaquette SE by Koga and Kawakami\cite{Koga99a}. Their result
for the triplet dispersion, is included by the dots in panel c) and is in
excellent agreement with our findings.

\section{Quantum Phase Transitions}\label{QPD}
In the following we consider the lines of quantum phase transitions
resulting from the closing of the elementary triplet-gap. This
provides for an upper bound on the extent of the region of stability
of the VBC in a tentative quantum phase diagram of the CCM.  In
principle additional first order transitions could occur or excited states
other than the elementary triplets could
collapse onto the ground state as well, leading to additional phase
boundaries. Here we focus on the triplet instability. Moreover, we
limit our discussion to AFM couplings $j_{1(2)}>0$ and
consider ferromagnetic cases elsewhere\cite{WBunpublished}.

Before proceeding we stress that the CCM is {\em symmetric}
under the interchange $j_1\leftrightarrow j_0$ and a diagonal shift by
one unit cell. This implies that the {\em complete} quantum phase diagram
can be obtained from a mirror reflection of the region $0<j_1<1$ and
$0<j_2$ at the line $(j_1=1,j_2)$ and a subsequent relabeling of
the axis by $j_1\rightarrow j_0/j_1\equiv 1/j_1$ and
$j_2\rightarrow j_2/j_1$.

\begin{figure}[tb]
\center{\psfig{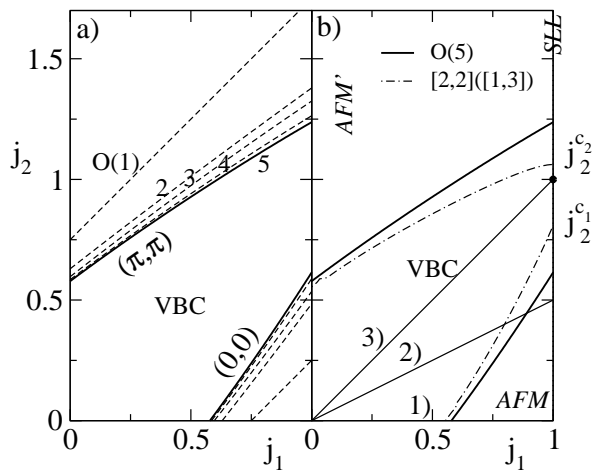}\hspace*{0cm}}
\caption{Triplet instabilities of the generalized CCM: a) Lines of
vanishing triplet-gap for increasing orders of the {\em bare} SE from
$O(1)$ to $O(5)$. Triplet softening occurs at the wave vector $(\pi,\pi)$
and $(0,0)$ for the upper and lower set of lines. VBC labels the region of
a finite triplet gap.  b) Comparison of the lines of vanishing triplet-gap
at $O(5)$ of the bare SE (solid) with the corresponding [2,2] and [1,3]
Dlog-Pad\'e approximants (dashed-dotted). The latter two are
indistinguishable on the scale of the plot. Along the selected directions
1), 2) and 3) cuts from the Dlog-Pad\'e analysis are be exemplified in
fig.~\ref{fig4}. The solid dot at $j_1$=$j_2$=1 refers the checkerboard
point.  The regions labeled by italic style {\it AFM}(') are likely to
exhibit antiferromagnetic
ordering\cite{Singh98a,Sindzingre02a,Singh99a,Koga99b,Voigt02a}. On the
line $(j_1=1,j_2)$ and for $j_2>j^{c_2}_2$ the ground state is a 'sliding
Luttinger liquid'\cite{Emery00a,Vishwanath01a,Starykh02a} ({\it SLL}). The
remainder of this diagram for all coupling constants is obtained by reflection
along $(j_1=1,j_2)$ and a rescaling of $j_1\rightarrow 1/j_1$ and
$j_2\rightarrow j_2/j_1$ (see text).}
\label{fig3}
\end{figure}

\begin{figure}[tb]
\center{\psfig{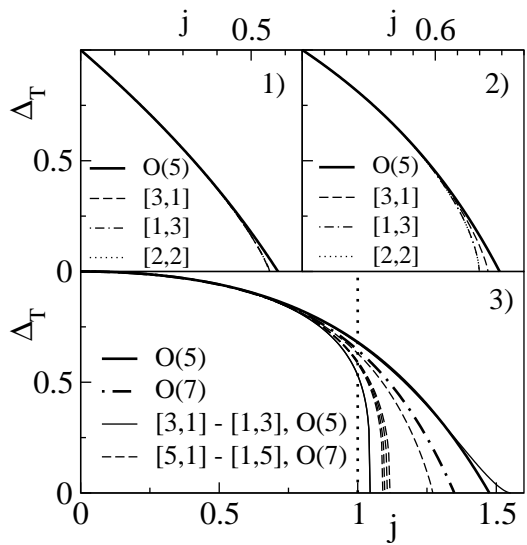}\hspace*{0cm}}
\caption{Spin gap $\Delta_{T}$ of the bare series (thick solid and
thick dashed-dotted) vs. reintegrated Dlog-Pad\'e
approximants along three selected directions 1), 2), and 3) shown in
fig.~\ref{fig3}b), i.e. $j_1=j$ and $j_2=aj$ with $a=0,0.5$ and $1$. The
$O(7)$ SE
in
panel 3) is from ref.\cite{Brenig02a}.  At
$O(5)$ all ( the [2,2] and [1,3] ) approximants are indistinguishable on
the scale of panel 1) ( 2) and 3) ). At $O(5)$, in panel 2) and 3) ($O(7)$,
in panel 3) ) the [3,1] ([5,1]) approximant is split off from the remaining ones
and is closer to the bare SE.}
\label{fig4}
\end{figure}

The results are summarized in figs.~\ref{fig3} and \ref{fig4} which
display information from the plain series as well as from a Dlog-Pad\'e
analysis. In fig.~\ref{fig3}a) we show the evolution with increasing order
of the lines of vanishing triplet gap of the {\em bare} series. First,
this figure further clarifies the location of the critical wave-vector ${\bf
k}_c$ as anticipated already in fig.~\ref{fig2}. Second the
panel is intended to prove a monotonous and smooth behavior of the bare
series within the range of interest. Based on the $O(5)$ SE,
fig.~\ref{fig3}b) shows the transition lines resulting from single-variable
Dlog-Pad\'es which have been obtained by parameterizing $j_{2(1)}$ of the
bare SE by $j_{2(1)}= a j_{1(2)}$ for $j_{2(1)}<j_{1(2)}$.
The figure displays a [2,2] and a [1,3] Dlog-Pad\'e. They are
indistinguishable on the scale of the plot. The lower(upper) critical
value of $j_2$ which they yield for the closing of the triplet gap on the line
($j_1$=1,$j_2$) are $j^{c_1}_2= 0.81$ and $j^{c_2}_2= 1.06$.
As discussed in the next paragraph, along the diagonal, i.e. for $j_1$=$j_2$,
Dlog-Pad\'es can be obtained also from an $O(7)$ SE\cite{Brenig02a}. They
result in somewhat larger values of the critical couplings than the $O(5)$ SE.
Roughly interpolating the difference in $j^c_2$ between the $O(5)$
and $O(7)$ Dlog-Pad\'es along this direction down to the line ($j_1$=1,$j_2$) we
approximate an error for the critical couplings of $j^{c_1}_2= 0.79 \ldots 0.81$
and $j^{c_2}_2= 1.06 \ldots 1.13$. These values should be
contrasted against LSW theory\cite{Singh98a}, which predicts stability of
an AFM phase for $j_2/j_1\leq 0.76$ at $j_1=1$. Moreover, ED at $j_1=1$
for up to 36 spins\cite{Sindzingre02a} suggests a closing of the spin-gap
below $j_2/j_1\approx 0.65$ and a crossover to 1D SLL-behavior at
$j_2/j_1\sim 1.5$.

The role of the Pad\'e analysis in fixing the location of the quantum
critical points as compared to the bare SE is clarified in more detail in
fig.~\ref{fig4} where several reintegrated Dlog-Pad\'es are displayed along
with the plain series for three selected directions shown in
fig.~\ref{fig3}b). For $j_1=j_2$, where the Pad\'e analysis is expected to be
most relevant, an $O(7)$ SE from ref.\cite{Brenig02a} has been included in
fig.~\ref{fig4}.3). Because of the symmetries of the quantum phase diagram
the curves beyond the vertical dashed line in fig.~\ref{fig4}.3) do not
refer to actual values of the spin gap for $j>1$. This range is shown for
completeness sake only. Most important, from fig.~\ref{fig4}.3) one
realizes that both, the plain SE as well as all of the Pad\'e approximants
yield {\em no} critical points for $j\leq 1$. While this agrees with
earlier findings of a finite spin gap at $j=1$ in
refs.\cite{Brenig02a,Sindzingre02a,Fouet03a} it seems that the checkerboard
point is almost critical. Panel 3) shows that the [4,2]-[1,5] and the
[2,2]-[1,3] approximants at $O(7)$ and $O(5)$, respectively, are well
clustered with critical points clearly separated from those of the
corresponding plain SE and the [5,1] and [3,1] approximants. Therefore, the
Dlog-Pad\'e analysis at $O(5)$ and arbitrary $j_1$, $j_2$ in
fig.~\ref{fig3}b) has been based on the [2,2] and [1,3] approximants. The
effects of the Dlog-Pad\'es in figs.~\ref{fig4}.1) and 2) is rather minute,
with the [2,2] and [1,3] approximants leading to essentially identical spin
gaps.

As noted already, direction 1) in fig.~\ref{fig3}b) refers to the
plaquette square-lattice Heisenberg model. Its transition to the AFM state
for $j\rightarrow 1$ has been studied by several groups.  We may therefore
compare the critical coupling $j_c=0.555$ which we find with that
obtained by other SE investigations, i.e. $j_c=0.555$ in ref.\cite{Singh99a}
and $j_c=0.54$ in ref.\cite{Koga99b}, as well as by ED, i.e. $j_c=0.6$ in
ref.\cite{Voigt02a}.  The agreement is satisfying.

Regarding the symmetry of the ground state in region AFM(') in
fig.~\ref{fig3}b), no conclusions can be drawn from the quadrumer SE.
Yet, subsuming the present work with
refs.\cite{Singh98a,Sindzingre02a,Singh99a,Koga99b,Voigt02a}
it is very likely that region AFM exhibits simple Ne\'el-type AFM-LRO.
Within AFM' the CCM
acquires a bipartite lattice-structure as $j_1\rightarrow 0$ which could
trigger a '4-up-4-down'-type AFM-LRO of the crossed chains for $j_2\gtrsim 1$.
The crossover between this and the SLL phase as $j_1\rightarrow 1$ is an open
issue.

\begin{figure}[tb]
\center{\psfig{file=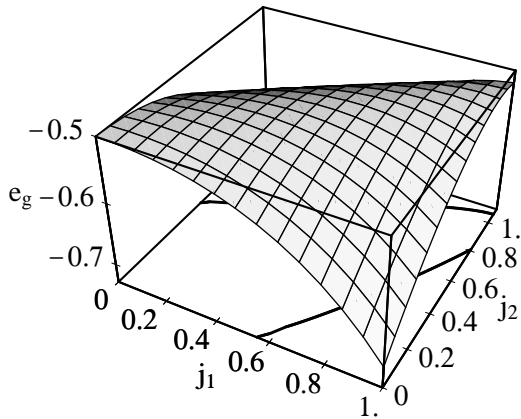,width=0.8\columnwidth}\hspace*{.5cm}}
\vspace{-.5cm}
\caption{$O(6)$ ground-state energy $e_g$ per spin from eqn.
(\ref{w4}). $e_g$ resembles the ground-state energy of the CCM only
within the range of stability of the VBC bounded by thick solid
lines in the ($j_1$,$j_2$)-plane, i.e. the dashed-dotted lines from
fig.~\ref{fig3}b).}
\label{fig5}
\end{figure}

\section{Ground State Energy}
To obtain SEs for the ground state energy valid to $O(n)$ in the
thermodynamic limit, i.e. for systems of infinite size one may evaluate the
matrix element $\langle 0|H_{\rm eff}|0\rangle$ on single 'maximal'
clusters, i.e. with periodic boundary conditions and sufficiently large not
to allow for wrap around at graph-length
$n$\cite{Knetter00,Brenig02a}. Here, $Q$-conservation of $H_{\rm eff}$ is
the main ingredient which reduces the size of the intermediate Hilbert
spaces. However, for the CCM we find the maximal-cluster approach to be
severely limited by memory-constraints and we have combined the
flow-equation approach with a linked cluster expansion. Some details of the
latter are commented on in appendix \ref{App1}. Up to $O(6)$ the ground
state energy $e_g$ {\em per spin} is
\begin{eqnarray}
\label{w4}
&& e_g = -\frac{1}{2}
-\frac{67 j_1^2}{576}
-\frac{481 j_1^3}{13824}
-\frac{17951 j_1^4}{663552}
-\frac{5705977 j_1^5}{522547200}
\nonumber\\ &&
-\frac{13033565594599 j_1^6}{2123765592883200}
+\frac{2 j_1 j_2}{9}
+\frac{59 j_1^2 j_2}{1728}
+\frac{224267 j_1^3 j_2}{4354560}
\nonumber\\ &&
+\frac{388714973 j_1^4 j_2}{35115171840}
+\frac{655905584767 j_1^5 j_2}{130026464870400}
-\frac{67 j_2^2}{576}
\nonumber\\ &&
+\frac{59 j_1 j_2^2}{1728}
-\frac{4837781 j_1^2 j_2^2}{104509440}
+\frac{75848383 j_1^3 j_2^2}{175575859200}
\nonumber\\ &&
-\frac{169095132323 j_1^4 j_2^2}{168552824832000}
-\frac{481 j_2^3}{13824}
+\frac{574241 j_1 j_2^3}{13063680}
\nonumber\\ &&
+\frac{38923349 j_1^2 j_2^3}{58525286400}
+\frac{113410023666229 j_1^3 j_2^3}{15928241946624000}
-\frac{139031 j_2^4}{5971968}
\nonumber\\ &&
+\frac{1184653457 j_1 j_2^4}{175575859200}
-\frac{65325840449533 j_1^2 j_2^4}{10618827964416000}
-\frac{2624063 j_2^5}{313528320}
\nonumber\\ &&
+\frac{181444615182347 j_1 j_2^5}{31856483893248000}
-\frac{51497150603363 j_2^6}{10618827964416000}
\end{eqnarray}
First we note, that at the checkerboard point, i.e. for $j_1=j_2$ eqn.
(\ref{w4}) is identical to the corresponding SE for $e_g$ in eqn. (6) of
ref.\cite{Brenig02a}. Figure~\ref{fig5} visualizes $e_g$,
resembling a convex function with the weakest energy gain in
the vicinity of the line of strongest frustration, i.e. for $j_1=j_2$. Note,
that fig.~\ref{fig5} is the ground state energy of the CCM only within the
region of stability of the VBC. Obviously $e_g$ is {\em below} the ground state
energy of the bare quadrumer limit for all $j_1$, $j_2$. This is
consistent with refs.\cite{Brenig02a,Sindzingre02a,Fouet03a} and is at variance
with a 'tetrahedral' SE to $O(3)$ at $j_1$=$j_2$ in
ref.\cite{Elhajal01a}, where $e_g > -0.5$ has been found.

\begin{figure}[bt]
\center{\psfig{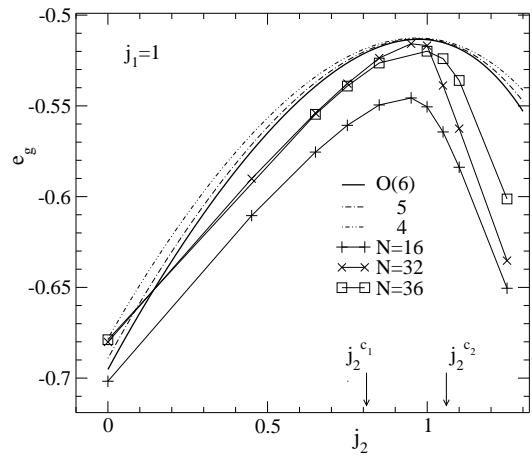}\hspace*{.5cm}}
\caption{Ground state energy of the CCM on the line ($j_1=1$,$j_2$).
Thick solid, dashed-dotted, and dashed-double-dotted lines: SE.
Solid curves with symbols: ED from fig. 2 of ref.\cite{Sindzingre02a}.
$j^{c_{1(2)}}_2$ critical couplings from Dlog-Pad\'e at $O(5)$,
see fig.~\ref{fig3}b).}
\label{fig6}
\end{figure}

In ref.\cite{Brenig02a} it has been shown, that $e_g$ at $j_1$=$j_2$ agrees to
within 1\% with that of ED on systems with 36 spins for $j_1\leq 1$.
In fig.~\ref{fig6} we compare $e_g$ with ED-results along the line
($j_1=1$,$j_2$) obtained by Sindzingre,
Fouet, and Lhuillier\cite{Sindzingre02a} for  systems with 16, 32, and 36 spins.
The SE is
displayed along with its variation in going from $O(4)$ to
$O(6)$ in order to provide a rough estimate of convergence. As with
fig.~\ref{fig5} the SE resembles the actual ground state energy
only within the range of critical couplings $j^{c_{1(2)}}_2$
marked on the $j_2$-axis.
While the line ($j_1$=1,$j_2$) is rather remote from
the limit of decoupled quadrumers, the agreement between SE and ED on the
largest system depicted in fig.~\ref{fig6} is still satisfying - even below
$j^{c_1}_2$.

Finally, one may define a 'point of maximum frustration', i.e. $j^{mf}_2$
at maximum $e_g$ for $j_1$=1. Interestingly
$j^{mf}_2$ $\neq 1$, i.e. it is {\em off} from the checkerboard point. We find
that $j^{mf}_2\approx 0.973\pm0.004$. The error has been approximated from
twice the difference between $j^{mf}_2$ at $O(5)$ and $O(6)$. As similar
observation has been reported in ref.\cite{Sindzingre02a}.

\section{Conclusion}\label{Conlus}
To summarize, we have investigated the ground state properties and the 
elementary triplet excitations of the generalized crossed-chain quantum-spin
model by quadrumer series expansion.
Over a large region of exchange parameters the model exhibits a finite spin gap
which is smoothly connected to the limit of the bare quadrumers. This region is
bounded by lines of triplet softening which results from a delocalization of the
triplets away from the case of complete frustration of the inter-quadrumer
exchange.
This remains true in particular not only at -- but also in the vicinity of the
checkerboard point for $j_1$=1 where we have established a finite range of
on-chain exchange, i.e. 0.79...0.81$\lesssim j_2\lesssim$1.06...1.13 with
a non-zero spin gap. While the fate of more complex multiparticle
excitations as a function of $j_{1(2)}$ remains to be investigated in the
future, our findings are consistent with a ground-state symmetry in the region
of non-zero spin gap identical to that at $j_{1(2)}$=0, i.e. a 
valence-bond-solid.

\indent {\bf Acknowledgments - } 
It is a pleasure to thank A. Honecker for fruitful discussions and D. C. Cabra
for helpful comments on SLLs. This research was supported in part by the
Deutsche Forschungsgemeinschaft, Schwer\-punkt\-programm 1073,
under Grant No.  BR 1084/2-3.

\begin{appendix}
\section{Linked Cluster Expansion}\label{App1}
\begin{figure}[bt]
\center{\psfig{file=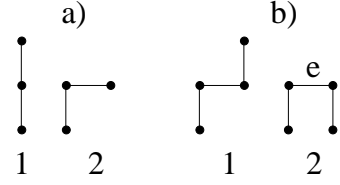,width=0.5\columnwidth}\hspace*{.5cm}}
\caption{Non-isomorphic (isomorphic) graphs of the quadrumer (standard)
expansion on the crossed-chain (simple square) lattice. Graph 2 is meant to
results from 1
by: a) an {\em in-plane} 90$^o$ rotation of the upper edge,  b) a
180$^o$-{\em out-off-plane} rotation of the upper edge around edge 'e'.}
\label{fig7}
\end{figure}

Here we comment on a peculiarity of graph counting for the quadrumer
expansion for the CCM as compared to that for standard 
expansions for spin lattice-models with no internal structure of the vertices
and edges. For the latter, both pairs of graphs in fig.~\ref{fig7}a) and b) are
isomorphic and yield identical results. In general
this is {\em not} the case for the quadrumer expansion. This is due to the
vertices(edges) to consist of
4-spin quadrumers(tetrahedral links) which prohibit 'free rotation' of a
graph at its vertices. Similarly, 'twisting' of an edge - as for edge 'e'
in fig.~\ref{fig7}b) - is {\em no} symmetry operation in general. In fact,
this depends on the choice of parameters, i.e. exactly at the checkerboard
point, for $j_1$=$j_2$, the tetrahedral link is symmetric under the
twist. This is reflected in, the corresponding cluster weights for the
ground state energies, which (discarding lattice constants) read
\begin{eqnarray}
\label{a1}
&&\mbox{Graph 1b)}=
-\frac{8597351609683 j_1^6}{6371296778649600}+
\nonumber \\ &&
\frac{9198603350203 j_1^5 j_2}{1327353495552000}
-\frac{473964630567761 j_1^4 j_2^2}{31856483893248000}+
\nonumber \\ &&
\frac{1700981391083 j_1^3 j_2^3}{94810963968000}
-\frac{143988201852347 j_1^2 j_2^4}{10618827964416000}+
\nonumber \\ &&
\frac{23692204192849 j_1 j_2^5}{3982060486656000}
-\frac{7166568732467 j_2^6}{6371296778649600}
\\[3pt]
&&\makebox{Graph 2b)}=
-\frac{7166568732467 j_1^6}{6371296778649600}+
\nonumber \\ &&
\frac{23692204192849 j_1^5 j_2}{3982060486656000}
-\frac{143988201852347 j_1^4 j_2^2}{10618827964416000}
\nonumber \\ &&
+\frac{1700981391083 j_1^3 j_2^3}{94810963968000}
-\frac{473964630567761 j_1^2 j_2^4}{31856483893248000}+
\nonumber \\ &&
\frac{9198603350203 j_1 j_2^5}{1327353495552000}
-\frac{8597351609683 j_2^6}{6371296778649600}
\label{a2}
\end{eqnarray}
Only if $j_1$=$j_2$ eqns. \ref{a1} and \ref{a2} are identical. In conclusion,
the number of non-isomorphic graphs in the quadrumer expansion
for the CCM is significantly larger than that in standard
expansions for 2D square-lattice models.

\section{Triplet Hopping Amplitudes}\label{appB}
In this appendix, and for completeness sake, we list the triplet dispersion
$E_T({\bf k})$ in the thermodynamic limit up to $O(j_1^m,j_2^n)$ with
$m+n\leq 5$ which can be written as
\begin{eqnarray}
\lefteqn{E_T({\bf k},j_1,j_2)=t_{00}+}
\nonumber \\ & &
t_{10}(\cos (k_x)+\cos (k_y))+
\nonumber \\ & &
t_{11}\cos (k_x) \cos (k_y)+
\nonumber \\ &&
t_{20}(\cos (2 k_x)+\cos (2 k_y))+
\nonumber \\ &&
t_{21}(\cos (2 k_x) \cos (k_y)+\cos (k_x) \cos (2 k_y))+
\nonumber \\ &&
t_{22}\cos (2 k_x) \cos (2 k_y)+
\nonumber \\ &&
t_{30}(\cos (3 k_x)+\cos (3 k_y))+
\nonumber \\ &&
t_{31}\cos (k_x) \cos (k_y) (\cos (2 k_x)+\cos (2 k_y)-1)+
\nonumber \\ &&
t_{32}(\cos (3 k_x) \cos (2 k_y)+\cos (2 k_x) \cos (3 k_y))+
\nonumber \\ &&
t_{40}(\cos (4 k_x)+\cos (4 k_y))+
\nonumber \\ &&
t_{41}(\cos (4 k_x) \cos (k_y)+\cos (k_x) \cos (4 k_y))+
\nonumber \\ &&
t_{50}(\cos (5 k_x)+\cos (5 k_y))
\label{b1}
\end{eqnarray}
where the hopping amplitudes $t_{00,...,50}$ are given by
\begin{eqnarray}
t_{00} && =
1+\frac{145 j_1^2}{432}+\frac{2771 j_1^3}{13824}+
\frac{9633109 j_1^4}{34836480}+
\\ & &
\frac{440341849993 j_1^5}{2106910310400}-
\frac{187 j_1 j_2}{216}-\frac{1033 j_1^2 j_2}{4608}-
\nonumber\\ & &
\frac{281352751 j_1^3 j_2}{313528320}-
\frac{173134656371 j_1^4 j_2}{300987187200}+
\frac{145 j_2^2}{432}-
\nonumber\\ & &
\frac{1033 j_1 j_2^2}{4608}+
\frac{30651767 j_1^2 j_2^2}{26127360}+
\frac{13833647581 j_1^3 j_2^2}{35115171840}+
\nonumber \\ & &
\frac{2771 j_2^3}{13824}-
\frac{90015077 j_1 j_2^3}{104509440}+
\frac{5030806153 j_1^2 j_2^3}{19508428800}+
\nonumber \\ & &
\frac{81044221 j_2^4}{313528320}-
\frac{50889983953 j_1 j_2^4}{100329062400}
+ \frac{411234777481 j_2^5}{2106910310400}
\nonumber
\end{eqnarray}

\begin{eqnarray}
t_{10} && =
(j_1-j_2)
\bigg(-\frac{1}{3}-
\frac{j_1}{18}+
\frac{29827{j_1^2}}{497664}+
\frac{1511209{j_1^3}}{39813120}-
\nonumber \\
& &
\frac{j_2}{18}-
\frac{37667 j_1 j_2}{248832}-
\frac{5805011 {j_1^2} j_2}{119439360}+
\frac{12673{j_2^2}}{165888}-
\nonumber \\
& &
\frac{1265911 j_1 {j_2^2}}{23887872}+
\frac{4947067 {j_2^3}}{119439360}
\bigg)+
\nonumber \\
& &
\frac{562865130343{j_1^5}}{6320730931200}-
\frac{4144974245353{j_1^4} j_2}{12641461862400}+
\nonumber \\
& &
\frac{2307417015943 {j_1^3}{j_2^2}}{4213820620800}-
\frac{485959984667{j_1^2}{j_2^3}}{842764124160}+
\nonumber \\
& &
\frac{1577116657219 j_1{j_2^4}}{4213820620800}-
\frac{672640186583{j_2^5}}{6320730931200}
\\ \nonumber \\
t_{11} && =
4 (j_1-j_2)\bigg(-\frac{j_1}{9}-\frac{7 j_1^2}{162}+
\frac{1221199 j_1^3}{89579520}+
\nonumber \\ &&
\frac{705849089 j_1^4}{25082265600}+
\frac{j_2}{9}-\frac{3836587 j_1^2 j_2}{29859840}-
\frac{23977237 j_1^3 j_2}{214990848}+
\nonumber \\ &&
\frac{7 j_2^2}{162}+
\frac{1199609 j_1 j_2^2}{9953280}+
\frac{20880367 j_1^2 j_2^2}{3135283200}-
\frac{507919 j_2^3}{89579520}+
\nonumber \\ &&
\frac{39361331 j_1 j_2^3}{358318080}-
\frac{2514549451 j_2^4}{75246796800}\bigg)
\\ \nonumber \\
t_{20} && =
2 {{(j_1-j_2)}^2} \bigg(-\frac{1}{54}-\frac{125 j_1}{5184}-
\frac{5789171 j_1^2}{179159040}-
\nonumber \\ & &
\frac{2583021851 j_1^3}{120394874880}-
\frac{125 j_2}{5184}+\frac{57485 j_1 j_2}{17915904}-
\nonumber \\ & &
\frac{176614487 j_1^2 j_2}{85996339200}-
\frac{5789171 j_2^2}{179159040}-
\frac{62829919 j_1 j_2^2}{9555148800}-
\nonumber \\ & &
\frac{11680356743 j_2^3}{601974374400}\bigg)
\\ \nonumber \\
t_{21} && =
4 {{(j_1-j_2)}^2} \bigg(-\frac{23 j_1}{486}-
\frac{875 j_1^2}{31104}-
\nonumber \\ & &
\frac{528075769 j_1^3}{56435097600}+
\frac{j_2}{18}-\frac{49 j_1 j_2}{116640}-
\frac{13388351 j_1^2 j_2}{391910400}+
\nonumber \\ & &
\frac{11 j_2^2}{384}+
\frac{3859528429 j_1 j_2^2}{112870195200}+
\frac{320776609 j_2^3}{37623398400}\bigg)
\end{eqnarray}

\phantom{to avoid wrong column breaks}

\begin{eqnarray}
t_{22} && =
4 {{(j_1-j_2)}^2} \bigg(-\frac{103 j_1^2}{2916}-
\frac{258601 j_1^3}{8398080}+
\nonumber \\ & &
\frac{317 j_1 j_2}{4374}+
\frac{1457333 j_1^2 j_2}{41990400}-\frac{293 j_2^2}{8748}+
\nonumber \\ & &
\frac{1207837 j_1 j_2^2}{41990400}-
\frac{266113 j_2^3}{8398080}\bigg)
\\ \nonumber \\
t_{30} && =
2 {{(j_1-j_2)}^2} \bigg(-\frac{j_1}{54}-
\frac{1849 j_1^2}{466560}-
\nonumber \\ &&
\frac{1692644279 j_1^3}{56435097600}+
\frac{j_2}{54}+\frac{9653478197 j_1^2 j_2}{112870195200}+
\nonumber \\ &&
\frac{1849 j_2^2}{466560}-
\frac{10100761397 j_1 j_2^2}{112870195200}+
\nonumber \\ &&
\frac{1870410679 j_2^3}{56435097600}\bigg)
\\ \nonumber \\
t_{31} && =
8 {{(j_1-j_2)}^3}
\bigg(-\frac{31 j_1}{1458}-\frac{838229 j_1^2}{41990400}+
\frac{j_2}{54}+
\nonumber \\ &&
\frac{7927 j_1 j_2}{3499200}+
\frac{871649 j_2^2}{41990400}\bigg)
\\ \nonumber \\
t_{32} && =
4 {{(j_1-j_2)}^2} \bigg(-\frac{2131 j_1^3}{78732}+
\frac{247 j_1^2 j_2}{2916}-
\nonumber \\ & &
\frac{2347 j_1 j_2^2}{26244}+
\frac{845 j_2^3}{26244}\bigg)
\\ \nonumber \\
t_{40} && =
2 {{(j_1-j_2)}^4} \bigg(-\frac{23}{17496}-
\frac{497953 j_1}{83980800}-
\nonumber \\ & &
\frac{497953 j_2}{83980800}\bigg)
\\ \nonumber \\
t_{41} && =
4 {{(j_1-j_2)}^4}
\bigg(-\frac{637 j_1}{52488}+
\frac{773 j_2}{52488}\bigg)
\\ \nonumber \\
t_{50} && = -\frac{161 {{(j_1-j_2)}^5}}{26244}
\label{b13}
\end{eqnarray}


\end{appendix}


\begin{thebibliography}{99}
\bibitem{HttpRef}
On apparence of this paper additional results are available on request from
the authors. These results are too lengthy to be displayed in written
form but will be made available electronically.
\end{thebibliography}
\end{document}